
\magnification=1200
\hoffset=1truecm
\vsize=7.5in
\hsize=5in
\pageno=0
\tolerance=10000

\def\eg{{\it e.g.}}
\def\ie{{\it i.e.}}
\def\etal{{\it et al.}}
\def\thru#1{\mathrel{\mathop{#1\!\!\!/}}}
\def\lsim{\mathrel{\rlap{\lower4pt\hbox{\hskip1pt$\sim$}}
    \raise1pt\hbox{$<$}}}         
\def\gsim{\mathrel{\rlap{\lower4pt\hbox{\hskip1pt$\sim$}}
    \raise1pt\hbox{$>$}}}         
\hfuzz=5pt
\baselineskip 12pt plus 2pt minus 2pt
\hfill DFTT 73/93

\hfill hep-ph/9311365
\medskip
\hfill  November 1993
\bigskip
\centerline{\bf AN INSTANTON-INDUCED CONTRIBUTION}
\centerline{\bf TO THE DECAY OF THE $\eta_c$ INTO $p$--$\bar p$}
\vskip 36pt\centerline{Mauro Anselmino$^{a,b}$ and Stefano
Forte$^{b}$\footnote*{Address after December 1, 1993: CERN, CH-1211
Gen\`eve 23, Switzerland}}
\vskip 12pt
\centerline{\it {}$^a$ Dipartimento di Fisica Teorica, Universit\`a di Torino}
\centerline{\it and}
\centerline{\it {}$^b$I.N.F.N., Sezione di Torino}
\centerline{\it via P.~Giuria 1, I-10125 Torino, Italy}
\vskip .8in
{\narrower\baselineskip 12pt
\centerline{\bf ABSTRACT}
\noindent We compute the decay rate for the process $\eta_c\to p\bar p$
using an effective helicity-flipping proton-antiproton-gluon vertex which
incorporates nonperturbative chiral symmetry breaking effects induced by
instantons. We fix the strength of the vertex by requiring it to account for
the screening of  the proton's axial charge observed in deep-inelastic
scattering,
and we estimate the size of the instanton effects by assuming
them to depend linearly on  the
instanton density. We find that, despite a large suppression, the
instanton-induced  process occurs with
a sizable  rate comparable to the observed one, whereas  the process is
forbidden in perturbative QCD and not understood using standard methods. }

\vskip 1.1in
\centerline{Submitted to: {\it Physics Letters B}}
\vfill
\eject
\input harvmac
The current theoretical understanding of perturbative QCD has improved at the
point that it is now possible to single out a few instances where the data
seem to disagree with perturbative computations. Polarization effects, in
particular, seem to display several cases where conventional perturbative
techniques appear to be inadequate.
At high energy, where perturbative methods should apply, polarized
phenomena are controlled by the helicity selection rules which follow from the
chiral symmetry of the QCD Lagrangian. However, as is well known
\ref\rom{See, \eg, R.~Jackiw, in S.~B.~Treiman, R.~Jackiw, B.~Zumino and
E.~Witten,``Current Algebra and Anomalies'' (World Scientific, Singapore,
1985)},
this symmetry is partly broken at the quantum level: flavor singlet chiral
symmetry, which is an exact symmetry of the QCD Lagrangian, is broken by
the axial anomaly. Besides
its standard perturbative consequences (such as the perturbative mixing of the
fermion singlet axial current with gluonic operators \ref\kodaira{J.~Kodaira,
Nucl. Phys. {\bf B165}, 129 (1980)}), this allows
nonperturbative symmetry-breaking effects, such as those required to solve the
U(1) problem of QCD \nref\uone{G.~'t~Hooft, Phys. Rep. {\bf 142}, 357
(1986)} \refs{\rom,\uone}.
This suggests that a
related symmetry-breaking mechanism may be at the origin of the observed
discrepancies with perturbative expectations. Explanation of
these effects may thus lead to a
better quantitative understanding of chiral symmetry breaking, a
nonperturbative phenomenon in QCD.

In previous work \nref\for{S.~Forte,  Phys. Lett.
{\bf B224}, 189 (1989);  Nucl. Phys.
{\bf B331}, 1 (1990)}\nref\forshu{S.~Forte and E.~V.~Shuryak,  Nucl. Phys.
{\bf B357}, 153 (1991)}\nref\anfor{M.~Anselmino and S.~Forte, Phys. Rev. Lett.
{\bf 71}, 223 (1993)} \refs{\for{--}\anfor}
we have suggested a specific scenario where these effects can
be understood: we have shown that instanton-induced interactions
\ref\tho{G.~'t~Hooft,  Phys. Rev. {\bf D14},
3432 (1976)} lead to
chirality breaking processes which may be described \anfor\
in terms of a family of effective pseudoscalar
nucleon--nucleon--$n$-gluon vertices which flip the nucleon chirality
in a way which depends on the momentum transfer at the nucleon vertex, but not
on the energy of the process. Even though the strength of the effective
coupling cannot be computed, general arguments allow to compute different
processes in terms of a single coupling. In particular, assuming \for\
this coupling
to be responsible for the observed discrepancy \ref\pspin{For a review,
see G.~Altarelli, in ``The Challenging
Questions'', proceedings of the 1989 Erice school, Plenum, New York
(1990). See also R.~L.~Jaffe and A.~Manohar,  Nucl. Phys., {\bf B337},
509 (1990)}
between the measured axial
charge of the nucleon and its quark model value allows  to compute the
single-spin polarization in elastic proton-proton scattering at high energy and
small momentum transfer, and predict that it satisfies a scaling law \anfor.

Here, we will apply the same physical mechanism to the decay of the $\eta_c$.
The decay of this particle into proton-antiproton poses an outstanding puzzle
to perturbative QCD \nref\diqa{M.~Anselmino, F.~Caruso, S.~Forte and
B.~Pire, Phys. Rev. {\bf D38}, 3516 (1988)}\nref\diqb {M.~Anselmino, F.~Caruso
and S.~Forte, Phys. Rev. {\bf D44}, 1438 (1991)}\nref\etarev{M.~Anselmino,
F.~Caruso and F.~Murgia, Phys. Rev. {\bf D42}, 3218 (1990)
}\refs{\diqa{--}\etarev}.
This particle is the lightest $c\bar c$ meson; its mass ($m_{\eta_c}=2980$~MeV)
excludes decays into particles with open charm and all of its decays
must proceed through gluon emission
and are Zweig
suppressed. The energy scale for these
processes is large enough that they ought to be mediated by hard gluons, as
required to explain their (observed) Zweig suppression, and in agreement with
the successful description of analogous charmonium decays.

Nevertheless, a
perturbative computation of the decay rate for $\eta_c\to p\bar p$
leads to a dramatic disagreement with the data: spin and parity
considerations
imply that the final $p\bar p$ state must be in a total spin $S=0$ state,
whereas, in
the limit of exact chiral symmetry, helicity conservation forces the final
$p\bar p$ state to have $S=1$, hence, the process is strictly forbidden. Chiral
symmetry breaking effects due to the quark masses can lead to a tiny decay
rate, of the order of few eV \ref\mass{M.~Anselmino, R.~Cancelliere and
F.~Murgia, Phys. Rev {\bf D46}, 5049 (1992)}.
However, the decay is experimentally observed to take
place with a width $\Gamma= (12.1 \pm 6.1)$~keV \ref\etawid{Particle Data
Group, M.~Aguilar-Benitez \etal, Phys. Rev. {\bf D45}, Part~2 (1992)}.
Even introducing effective diquark
constituents in the nucleon wave function does not help: although these
would allow the decay without need for chiral symmetry violation (because of
the presence of spin-1 constituents) the computed rates are typically too
small by four orders of magnitude \diqb. An explanation of the large
observed $\eta_c \to
p\bar p$ decay rate seems to require more fundamental
modifications of the usual perturbative QCD: for example,
a nonperturbative gluonic coupling of
the $\eta_c$ to  the nucleon could provide such new mechanism
\nref\gluea{M.~Anselmino, M.~Genovese and E.~Predazzi, Phys. Rev. {\bf D44},
1597 (1991)}\nref\glueb{M.~Anselmino, M.~Genovese and D.E.~Kharzeev, Torino
preprint DFTT 64/93 (1993)}\refs{\gluea,\glueb}.

The effective instanton-induced nucleon--nucleon--$n$-gluon interaction
discussed in Ref.~\anfor\ incorporates nonperturbative chiral symmetry
breaking and could thus provide a dynamical model for
this kind of process. However, the effect discussed in
\anfor\ persists at arbitrarily high energies {\it provided} the limit
of vanishing momentum transfer at the nucleon vertex is taken; here, instead,
the momentum transfer is equal to the mass of the $\eta_c$. Furthermore, the
effective action of Ref.~\anfor\ has been derived in the semiclassical
approximation, which is justified if the strong coupling is
large, but it certainly fails in the asymptotic $\alpha_s\to 0$ limit;
also, in this limit any instanton effect should be exponentially suppressed.
Here we shall discuss these problems, and estimate the modifications to the
quantitative consequences of the results of Ref.~\anfor\ due to these effects.
We will see that the $\eta_c$ mass falls within an intermediate energy
range where the strong coupling
$\alpha_s$ is small enough that
a perturbative treatment is allowed, but (essentially
because of its slow logarithmic fall-off)
nonperturbative effects are still sizeable and observable in channels where
conservation laws forbid the bulk of the usual perturbative process. Even
though large theoretical uncertainties are involved in an estimate of instanton
effects, we will be able to show that their size is significantly
larger than any previously computed perturbative contribution and comparable to
the experimentally observed  rate.

Let us first review the derivation of the effective nucleon--nucleon--$n$-gluon
Lagrangian which we shall use \anfor. This is based on the observation \for\
that instantons may contribute to the axial form factor $G_A(q^2)$ of the
nucleon, defined by the decomposition
\eqn\ffdef
{\langle p^\prime,\lambda^\prime|j^\mu_5|p,\lambda\rangle=
G_A(q^2) \bar u_{\lambda^\prime}(p^\prime)\gamma^\mu\gamma_5u_\lambda(p)+
G_P(q^2)q^\mu\bar u_{\lambda^\prime}(p^\prime)\gamma_5u_\lambda(p),}
where $G_A$ and $G_P$ are respectively the axial and pseudoscalar form factors,
$u_\lambda(p)$ is a nucleon spinor with momentum $p$, mass $m_N$ and helicity
$\lambda$, and $q=p^\prime-p$ for space-like processes and $p^\prime+p$
for time-like ones. That the presence of a classical instanton background
field may contribute directly to $G_A(q^2)$ can be understood by
considering the time component of Eq.~\ffdef\ in the forward direction,
\ie,
\eqn\axch
{\langle p,\lambda|Q_5|p,\lambda\rangle=2\lambda \, G_A(0) .}
Eq.\axch\ shows that
$G_A(0)$ is the coefficient of proportionality between the nucleon's total
axial charge and its helicity: in the presence of instantons, the axial charge
receives \for\ a contribution which is not conserved and corresponds to
anomalous violation of chirality. An explicit mechanism through which such a
contribution may arise has been presented in Ref.~\forshu, in a simplified
model
(QCD with a single massless quark flavor and gauge group SU(2), and with
the nucleon matrix elements in Eq.\axch\ replaced by a quark matrix element).

Whereas this kind of contribution can be computed exactly only in simplified
models, such as that of Ref.~\forshu, its size may be estimated by {\it
assuming} that it is mainly responsible for the fact that the experimentally
measured value of $G_A(0)$ is very small, $G_A(0)\sim 0$, whereas the quark
model would lead to expect $G_A(0)\sim 0.6$. Such an  assumption is supported
by the model computation of Ref.~\forshu. This determines the value of the
instanton contribution to $G_A(0)$. The latter, however, may be viewed as
the strength of an
effective instanton-nucleon-nucleon coupling. Indeed, because of the anomaly
equation \rom, Eq.~\ffdef\ implies
\eqn
\flip
{\eqalign{\lim_{q\to 0}
iG_A(q^2)q_\mu \bar u_{\lambda^\prime}(&p^\prime)
\gamma^\mu\gamma_5u_{\lambda}(p)=\cr
&=\langle p,\lambda^\prime|
\left(-2Q + \sum_{\rm flavors}
2 i m_i \bar \psi_i\gamma_5 \psi_i\right)|p,\lambda\rangle,\cr}}
where $Q$ equals the number density
of instantons minus anti-instantons:
\eqn\idens
{Q=-{N_f\over 32 \pi^2} g^2 \tr
\epsilon^{\mu\nu\rho\sigma}F_{\mu\nu}F_{\rho\sigma}.}

Hence, the instanton contribution to the first term on the r.h.s. of
Eq.\flip\ provides (in the semiclassical approximation) an effective
instanton-nucleon-nucleon coupling, which we may rewrite in the form
\eqn\effcoup
{\langle p^\prime,\lambda^\prime|Q^{\rm Inst}|p,\lambda\rangle=-
im_NG_A^{\rm Inst}(q^2)\bar u_{\lambda^\prime}(p^\prime)\gamma_5
u_{\lambda}(p),}
where $G_A^{\rm Inst}(q^2)$ depends only on $q^2$
(so that $G_A^{\rm Inst}(0)$ is a universal coupling) \anfor.
Furthermore, in the semiclassical limit, the instanton can fragment into $n$
gluons (the semiclassical approximation being allowed if $n\gsim {1\over
\alpha_s}$), thereby leading to an effective  nucleon--nucleon coupling
with the production of $n$ gluons, described by the matrix element
\nref\blviol{A.~Ringwald,  Nucl. Phys {\bf B330}, 1 (1990); for a review see
M.~Mattis,  Phys. Rep., {\bf B214}, 159 (1992)}\refs{\blviol,\anfor}
\eqn\effglu
{\eqalign{\langle p^\prime,
\lambda^\prime|\prod_{i=1}^n\left[ A_{\nu_i}^{a_i}(k_i)
\right]&
|p,\lambda\rangle=(2\lambda)
im_NG_A^{\rm Inst}(q^2)\bar u_{\lambda^\prime}(p^\prime)\gamma_5
u_{\lambda}(p)\cr
&\times\prod_{i=1}^n\left({16\pi^2\over g_s}\right)\left[
{\tilde\eta^{a_i}_{\mu_i\nu_i}
k_i^{\mu_i}\over k_i^4}\left(1-{1\over2} K_2(\rho |k_i|) \rho^2 k_i^2
\right)\right],\cr}}
where $K_2$ denotes  the modified Bessel function,
$g_s$ is the strong coupling (\ie, $\alpha_s={g_s^2\over 4\pi}$),
$\rho$ is the instanton radius, which must eventually be integrated over,
$a_i$, $k_i^{\mu_i}$, and $\nu_i$ are respectively the
color\footnote*{The indices $a_i$ in Eq.\effglu\
run over an  SU(2) subgroup of the color gauge group, which
must be embedded into SU(3).}, four-momenta,
and Lorentz indices of the $n$ gluons (over which the product runs),
and $\tilde\eta^{a_i}_{\mu_i\nu_i}$ denotes the 't~Hooft
symbols \tho\ $\eta^{a_i}_{\mu_i\nu_i}$
or $\bar\eta^{a_i}_{\mu_i\nu_i}$ of the instanton or  anti-instanton
(which must also be summed over), respectively.

In Ref.~\anfor\ Eq.\effglu\ was used to derive an effective coupling in
the limit of small $k_i$ by the LSZ procedure; with $
|G^{\rm Inst}_A(q^2)|\sim |G^{\rm Inst}_A(0)|\sim 1$ from the
requirement that the instanton contributions to $G_A(0)$ cancels the quark
model one in order to lead to the near-vanishing experimental value, and $\rho$
fixed as the average instanton radius. In such case the strong coupling is of
order $\alpha_s\sim 1$ and the process is semiclassical with a small number of
gluons. Here we would like to use the coupling
Eq.~\effglu\ to  describe the production of a gluon state which couples
to the $\eta_c$, so that its decay may proceed through the diagram of
Fig.1.
To this purpose, we must make sure on the following points: first, we should
establish that the process of Fig.~1 exists, in the sense that semiclassical
gluons may couple perturbatively to the $c$ quark line; second,
we should determine how the size
of the instanton effects is affected by the the extrapolation from $q=0$ to
$q\sim m_{\eta_c}$.

The existence of the process of Fig.~1 relies on the slow fall-off of
$\alpha_s$.  At $m_c$ the value of the strong coupling is known with
good accuracy from spectroscopy \ref\alphach{W.~Kwong, P.B.~Mackenzie,
R.~Rosenfeld and J.L.~Rosner, Phys. Rev. {\bf D37}, 3210 (1988)},
and it is given by $\alpha_s(m_c)=0.28$. This value is
small enough that the coupling of the gluons to the charm quark may be treated
perturbatively, as is borne out by ample spectroscopic evidence.
The coupling of gluons emitted from the charmonium
decay to the nucleon, however, will in general be nonperturbative.
Because ${1\over \alpha_s}\sim 3$, instanton-induced
processes with few gluons are
semiclassical at this scale. Hence gluons radiated from the $\eta_c$ decay may
couple in principle to the semiclassical vertex of Eq.~\effglu. Because
the $\eta_c$ is $C$-even the leading contribution will be given by the
two-gluon process of Fig.~1. Whether this is quantitatively significant is
the question we address next.

Quite in general, one would expect that instanton effects are weighted by the
Euclidean instanton action $e^{-{2\pi\over\alpha_s}}$; thus even the moderate
decrease in coupling from $\alpha_s\sim 1 $ to $\alpha_s\sim{1\over3}$
yields a substantial suppression. In order to estimate this suppression
quantitatively, recall that
the instanton contribution to $G_A(q)$ \forshu\ is due to the fact
that instantons behave in a dielectric way with respect to the axial charge,
\ie, in presence of a source carrying axial charge they give rise to anomalous
creation of axial charge anticorrelated to that of the source; thus,
the effect
will be weighted by the probability of interaction with an instanton.
We will crudely estimate this probability by assuming it to be
linear in the instanton density.

Now, because the effective coupling Eq.~\effcoup\ is due to anomalous
axial charge creation in the instanton background, it will only receive
contributions from instantons of size $\rho\lsim{1\over q}$, because the
anomalous particle creation induced by the instanton occurs at a finite
distance of order\footnote*{For example, it can be proven
rigorously \ref\nico{R.~Guida, K.~Konishi and N.~Magnoli, Genova preprint
GEF-Th-14/93 (1993)}
that particle creation in an instanton--anti-instanton valley background
disappears if the separation of the instanton--anti-instanton pair is smaller
than ${4\over3}\rho$.}
$\rho$. It follows that no particle creation is seen if an instanton of radius
$\rho$ is probed at a scale $q> {1\over \rho}$; consequently instantons
with $\rho>\rho_c$, with $\rho_c\sim{1\over q}$,
do not contribute to the effective interaction Eq.~\effcoup\
and its cognates Eq.~\effglu. Accordingly, the instanton density, on which we
assumed $G_A^{\rm Inst}$ to depend linearly, will decrease from its vacuum
value $n_0$ (which includes instantons of all sizes) to its value $n(\rho_c)$,
computed including only small
instantons with $\rho<\rho_c$. Thus we estimate
\eqn\suppr
{N(q)\equiv{G_A^{\rm Inst}(q)\over G_A^{\rm Inst}(0)}\sim {n({1/ q})\over
n_0}.}

The suppression due to the decrease in instanton density can now  be
estimated
using Eq.~\suppr\ and the explicit expression of the differential instanton
density\footnote*{We use here the {\it vacuum} instanton density. In
general, the quark-quark chirality flipping interaction \tho\
will carry extra powers of $\rho$; these, however, are  cancelled by
corresponding powers of $\rho$ in the quark propagators in the instanton
background when computing a matrix element, such as that of Eq.~\effcoup, as
it is clear on dimensional grounds.} \nref\shifden{M.~A.~Shifman,
A.~I.~Vainshtein and V.~I.~Zakharov,
Phys. Lett. {\bf 76B}, 471}\nref\shu{See E.~V.~Shuryak, Phys. Rep. {\bf 115},
151 (1984)}\refs{\tho,\shifden,\shu}
\eqn\iden
{{dn(\rho)\over d\rho}= {C\over \rho^5} \left[\alpha_s(\rho^{-1})\right]^{-6}
\exp\left[-{2\pi\over\alpha_s(\rho^{-1})}\right]}
where all the $\rho$--independent quantities have been lumped in the constant
$C$. The differential density Eq.\iden\ should then
be integrated over all instanton
radii. The density of instantons with
radii up to $\rho_c$ will be given integrating in the range $0\leq \rho\leq
\rho_c$; because in the case of interest $\rho_c$ is in the perturbative
region the integration can be performed by using the perturbative expression
for $\alpha_s$. The vacuum instanton density in the  denominator of Eq.~\suppr,
instead, of course
diverges for large instanton radii; this problem is cured phenomenologically
\shifden\ by assuming the growth of the instanton radius to be cut off at some
scale $\rho_0\sim 1$~Fm by a (poorly known) instanton repulsion mechanism.
The corresponding value of the average instanton radius (obtained averaging
with the measure Eq.~\iden\ up to $\rho_0$) is \shu\ of order of
$\langle\rho\rangle\approx {1\over 3}$~Fm.

In order to minimize the model dependence of the computation of $N(q)$,
Eq.~\suppr, rather than using a phenomenological value for $n_0$ we compute
the denominator of Eq.~\suppr\ by integrating Eq.~\iden\ up to a cutoff
value $\rho_0\sim 1$~Fm. This indeed minimizes the model-dependence of the
result because then the value of $N(q)$ is controlled by the
exponential dependence of
$n(\rho)$ on $\rho$ around $\rho\sim m_{\eta_c}^{-1}$,
while it is essentially
independent of pre-exponential factors. The former
in turn is
controlled by
the value of $\alpha_s(m_{\eta_c})$, which is known rather accurately.
On the contrary, as we will shortly see, the result
is rather insensitive to both the upper limit of
integration $\rho_0$, and the precise value of $\alpha_s(\rho_0^{-1})$, which
is large, $\alpha_s(\rho_0^{-1})\gsim 1$.

Due to sensitivity of the results to the perturbative running coupling, we
use the next-to-leading
form
\eqn\alphas
{\alpha_s(Q^2)={4\pi \over \beta_0 \ln t}
\left[ 1- {\beta_1 \ln \ln t \over \beta_0^2 \ln t} \right] ,}
where $\beta_0 = 11 - 2n_f/3$, $\beta_1 = 102 - 38n_f/3$
and in the perturbative regime
$t = \ln (Q^2/\Lambda^2)$.
In order to compute the suppression Eq.~\suppr, we must introduce an infrared
interpolation for the strong coupling $\alpha_s$, so that at large $Q^2$ the
perturbative behavior is reproduced, while at small $Q^2$  the
strong coupling $\alpha_s$ saturates
to a constant value. To this purpose we set
$t = \delta + \ln (Q^2/\Lambda^2)$.
For the perturbative QCD scale we take the  value
(with $n_f=4$) $\Lambda = 263$ \ref\lqcd{M.~Virchaux and
A.~Milsztajn, Phys. Lett. {\bf B274}, 221 (1992)},
which yields $\alpha_s(m_c)=0.31$, in good agreement with spectroscopy,
whereas  $\delta$ is fixed by
the value of $\alpha_s$ in the infrared.
For example, imposing \shifden\ $\alpha_s(1\>{\rm Fm}^{-1})=1$
 we get $\delta=3$,
while
larger infrared values of $\alpha_s$ can be
obtained by reducing the value of $\delta$
(which must anyway satisfy $\delta>1$ to insure saturation).

With this form of the strong
coupling\footnote*{The computation of the integral over instanton
radii has been performed by taking the appropriate number of flavors
at any scale, \ie, $n_f=3$ for ${1\over \rho}<m_c$, $n_f=4$ for
$m_c\le {1\over\rho}<m_b$ and so forth, and suitably updating the value of
$\Lambda$ \ref\mar{W.~Marciano, Phys. Rev. {\bf D29}, 580 (1984)}. The result
turns out to be essentially the same as that obtained setting $n_f=4$, due
to the rapid falloff of the instanton density at large $\rho$, as well as the
insensitivity of  the value of $N(q)$ Eq.~\suppr\ to the precise form of the
coupling in the infrared.}, $\delta=3$,
and taking the cutoff radius of the vacuum instanton density to be
$\rho_0=1$~Fm, we get $N^2(m_{\eta_c})\simeq 3.9 \times 10^{-5}$.
Varying the infrared value of $\alpha_s$ by an order of magnitude as well
as the value of $\rho_0$ from 1~Fm to $10$~Fm this determination varies at most
by a factor two, thus displaying the infrared stability of our estimate.
Even though there is (as expected) a substantial suppression
of the coupling Eq.~\effglu\ due to the decrease in instanton density, the
strength of the coupling is still large enough to lead to a sizable decay rate,
as we show next.

We proceed therefore to the computation of the diagram of Fig.~1.
The coupling of gluons to the charm quark line is assumed to be given by
perturbative QCD; the charm
quark-antiquark pair then hadronizes with a wave function
that we assume to have the static (nonrelativistic) form,
\nref\barb{R.~Barbieri,
R.~Gatto and R.~Kogerler, Phys. Lett {\bf 60B}, 183 (1976)} \refs{\barb,\diqb}
\eqn\etawf
{\psi_{\eta_c}(\vec k)={1\over 4\sqrt 3} \, R(0) \, {\delta (|\vec k|)\over
(\vec k)^2} \, \delta_{ij} \, (c_+^i \bar c_+^j - c_-^i \bar c_-^j),}
where $i,j$ are colour indices and $\pm$ denote the quark helicities.
The value of the radial charmonium wave function in the origin, $R(0)$, can
be fixed  computing \barb\ the width of the electromagnetic process
$\eta_c\to\gamma\gamma$ and comparing to the experimental value \etawid:
\eqn\rzero
{\Gamma(\eta_c \to \gamma\gamma)={64\over 27} {\alpha^2 \over m_{\eta_c}^2}
\vert R(0) \vert^2 \left( 1-3.4{\alpha_s\over \pi}\right) =
(6.6^{+2.4}_{-2.1})\>{\rm keV},}
which yields $R^2(0) = (0.67 ^{+0.24}_{-0.21})$~(GeV)$^3$.

The amplitude for the decay process is then given by
\eqn\ampl
{A_{\lambda_p,\,\lambda_{\bar p}}=\pi R(0)
\left[ M_{\lambda_p,\,\lambda_{\bar p};\>++}(\vec k=0) -
M_{\lambda_p,\,\lambda_{\bar p};\>--}(\vec k=0) \right],}
where $M$ are the elementary helicity amplitudes for the process
$c\bar c \to p\bar p$:
\eqn\amplb{
\eqalign{
M_{\lambda_p,\,\lambda_{\bar p};\>\lambda_c,\,\lambda_{\bar c}}
&=\int\,{d^4k_1\over (2\pi)^4}\,{d^4k_2\over
(2\pi)^4} \cr
\times & R_{\lambda_c,\,\lambda_{\bar c}}^{\mu\nu}(c,k) \,
T^{\lambda_p,\,\lambda_{\bar p}}_{\mu\nu}(k_1,k_2,p) \,
(2\pi)^4 \delta^{(4)}(k_1+k_2-2c), \cr}}
and we have set $c=(m_c,\vec 0)$, $m_c=m_{\eta_c}/2$ and $k=(k_1-k_2)/2$,
while  $R^{\mu\nu}$ and $T^{\mu\nu}$ give respectively the perturbative part of
the amplitude, and the nonperturbative coupling to the proton.
Explicitly, the perturbative contribution is given by
\eqn\pertam
{R_{\lambda_c,\,\lambda_{\bar c}}^{\mu\nu}(c,k)=
\bar v_{\lambda_{\bar c}}(c)\gamma^\mu{\thru k + m\over k^2 +
m^2}\gamma^\nu u_{\lambda_c}(c),}
where we have Wick-rotated to Euclidean space where the instanton coupling is
defined. The nonperturbative coupling is constructed by taking $n=2$ in
Eq.~\effglu\ and calculating the color trace, which, due to the
form of the wave function Eq.~\etawf\ amounts basically to
projecting out the color singlet component, with the result
\eqn\npertam
{\eqalign{T_{\lambda_p,\,\lambda_{\bar p}}^{\mu\nu}(&k_1,k_2,p)=
{(2\pi)^4\over 3g_s^2}\,G_A(m_{\eta_c})\>m_N \, 2\lambda_p \,
\bar u_{\lambda_p}(p)i\gamma_5v_{\lambda_{\bar p}}(p^\prime)\cr
&\times
\left[k_1\cdot k_2g^{\mu\nu}- {1\over 2}(k_1^\mu k_2^\nu+k_1^\nu k_2^\mu)
+(2\lambda_p)\epsilon^{\alpha\mu\beta\nu} k_{1\alpha} k_{2\beta}\right]
{\Phi(k_1)\Phi(k_2) \over k_1^4 k_2^4}.\cr}}
In Eq.~\npertam\ we have set $p^\prime=k_1+k_2-p$, and
\eqn\phidef
{\Phi(k)=
4\left(1-{1\over2} K_2(\rho |k|) \rho^2 k^2\right).}

Using Eq.s~\pertam\ and \npertam\ in Eq.~\amplb\ we get, through a somewhat
tedious but straightforward computation
\eqn\finres
{\eqalign{M_{\lambda_p,\,\lambda_{\bar p}\>\lambda_c,\,\lambda_{\bar c}}
&={(2\pi)^4\over 3g_s^2} G_A(m_{\eta_c}) \, 4m_c^2 \, (4\lambda_c \lambda_p)
\, \delta_{\lambda_c \lambda_{\bar c}}\delta_{\lambda_p \lambda_{\bar p}}
\cr &\times \int\,{d^4 k\over (2\pi)^4} {g_s^2 \over
k^4(2c-k)^2\left[(c-k)^2+m_c^2\right]}\Phi(k)\Phi(2c-k).\cr}}
The loop integration in Eq.~\finres\ converges, as one would expect due to the
effective nature of the interaction \effglu, because the function $\Phi(k)$,
Eq.~\phidef, behaves as
\eqn\phibeh
{\eqalign{\phi(k)&{\mathop \sim\limits_{k\to 0}} \rho^2 k^2 \cr
\phi(k)&{\mathop \sim\limits_{k\to \infty}}
4\left(1-{1\over2}\rho^2k^2\sqrt{\pi\over2\rho|k|}e^{-\rho|k|}\right)\cr}}
thereby cutting off the ultraviolet tail of the integration while being finite
in the infrared.
Using the result Eq.~\finres\ in the
expression \ampl\ of the decay amplitude
and supplementing the required
kinematical factors
the width for the  process is found to be
\eqn\width
{\Gamma(\eta_c\to\bar p p)={ m_N(m_c^2-m_N^2)^{1/2}\over 16 \pi^2}
\sum_{\lambda_p,\>\lambda_{\bar p}}|A_{\lambda_p,\>\lambda_{\bar p}}|^2.}

In order to determine the value of $\Gamma$, Eq.~\width, we must still
fix the values of the instanton radius and of the coupling constant in the
nonperturbative
portion of the amplitude, Eq.~\npertam. In principle, all $\rho$-dependent
quantities ought to be included in the averaging over instanton radii
which also provides the decrease of $G_A^{\rm Inst}$ according to Eq.~\suppr;
this, however, would unnecessarily (to the level of accuracy of this estimate)
complicate the evaluation of the loop
integral in Eq.~\finres. Hence, we evaluated instead
the integral by fixing
$\rho={1\over
m_{\eta_c}}$. Using the value quoted above of $N(m_{\eta_c})$ we get
thus finally
\eqn\resgamma{\Gamma \sim 0.8 \>{\rm keV}}

We can get a more direct handle on the ingredients which enter
the estimate Eq.~\resgamma\ by noting
that, because of the behavior
Eq.~\phibeh\ of the function $\Phi(k)$, the loop integration is effectively
cut off in the ultraviolet at momenta of order $|k|\sim {1\over \rho}$.
If accordingly we
neglect the contribution to the integral from  momenta
$|k|> {1\over \rho}$, using
the low-momentum form of $\Phi$, Eq.~\phibeh, the integration may be performed
analytically, with the result
\eqn\resan
{\Gamma = K \times |I|^2 \times |N|^2 \times {g_s^4(m_c)
\over g_s^4(\rho^{-1})},}
where all kinematical factors have been lumped in
\eqn\kinfac
{K = {m_N m_{\eta_c}^4 \sqrt{m_{\eta_c}^2-4m_N^2} \over 3\pi^2}
\vert R(0) \vert^2,}
while $I$ is the result of the loop integral,
\eqn\lint{\eqalign{
I &= \int\, d^4k {1 \over
k^4(2c-k)^2\left[(c-k)^2+m_c^2\right]}\Phi(k)\Phi(2c-k) \cr
&\approx  \rho^4 \int\, d^4k {1 \over
(2c-k)^2 \left[(c-k)^2+m_c^2\right]} \cr
& = \rho^4 \pi^2 \ln
{m_c^2 + {1\over \rho^2}\over m_c^2 \>}\cr}}
and $N$ provides the overall suppression due to
the instanton density, Eq.\suppr.
Numerically, $K \simeq 0.3$ (GeV)$^9$, while
(with $\rho^{-1}=m_{\eta_c}$)
$I \simeq 0.2$ (GeV)$^{-4}$
 and $N^2 \simeq 3.9 \times 10^{-5}$,
which indeed reproduces the result Eq.~\resgamma.

It thus appears that the decay rate induced by the chirality-flipping
interaction is kinematically of order of several hundred MeV, but it
is then reduced due to the large suppression of instanton effects
to the order of the keV, in good agreement with the experimental
value.
Clearly, this  is a crude estimate of the order of magnitude of
$\Gamma(\eta_c \to p\bar p)$ induced by instantons. The main
uncertainties involved in this estimate enter in the determination of
the suppression factor $N(q)$ Eq.~\suppr, mostly in the exact dependence
on the cutoff instanton radius and in the precise value of the strong
coupling. For example, if the value $\rho_c= {3\over 2}{1\over q}$ were
used (appropriate for the instanton--anti-instanton valley of
Ref.~\nico), rather than $\rho_c={1\over q}$, then the value of $N(q)$ would
increase by an order of magnitude. Analogously, varying  $\Lambda$
between 200 and 300 MeV the value of $N(q)$ fluctuates by about one
order of magnitude.
These uncertainties also affect the effective value of $\rho$ which should
be used in the computation of the loop integral Eq.~\finres, on which
the loop integral depends as $I\propto \rho^4$, according to Eq.~\lint.
As discussed above, a smaller uncertainty  comes from the infrared
behavior of the instanton density.
Finally, a minor uncertainty enters in the experimental determination
of $R(0)$.
In short, most of the uncertainty comes from the perturbative behavior
close to the mass of the $\eta_c$, which is under theoretical control
and may be reduced by a more accurate treatment and a better
determination of $\Lambda$. The result turns out to be in remarkable
qualitative agreement with experiment.

The instanton-induced coupling discussed here leads also to decay
of the first radial excitation of the $\eta_c$ according to the same mechanism,
\ie, to the process
$\eta^{\prime}_c \to p\bar p$.
This has not been observed yet,  even though
data are expected to become available soon.
If these processes proceed mainly through the instanton-induced
mechanism discussed here, we expect the value of
$\Gamma(\eta^{\prime}_c \to p\bar p)$ to be smaller than that for the
$\eta_c$.
Indeed, even though the larger value of the $\eta^{\prime}_c$ mass
($m_{\eta^\prime_c}\simeq 3.6$~MeV) leads to
an increase of  the phase space (compare
Eq.~\kinfac), this is largely compensated by the very rapid decrease of the
instanton suppression factor Eq.~\suppr\ (and to a lesser extent,
the loop integral Eq.~\lint) as  the characteristic scale of
the process increases.
Specifically, we get
\eqn\etap{
S \equiv
{\Gamma(\eta_c^\prime \to p\bar p) \over \Gamma(\eta_c \to p\bar p)}
\simeq 0.2 \times {|R(0)(\eta_c^\prime)|^2 \over |R(0)(\eta_c)|^2}}
The precise value of $S$ depends on
the ratio of the radial wave functions in
the origin. Assuming for instance this ratio to behave as
that of successive $s$-wave radial
excitations in a hydrogen-like potential,
then $R_{20}(0)/R_{10}(0) = 1/\sqrt 8$, leading to
$S \simeq 3 \times 10^{-2}$.

In sum, we have estimated the magnitude for the width of the decay
$\eta_c\to p\bar p$ using an instanton-induced effective Lagrangian to model
the
chiral symmetry breaking which is necessary in order for the decay to take
place. The decay proceeds through the emission by this effective Lagrangian
of a gluon pair which is soft enough to be treated semiclassically at the
proton vertex, but hard enough to couple perturbatively to a charmed quark
line.
The result turns out to be remarkably of the correct order of
magnitude, whereas all previously known mechanism to describe exclusive
processes within QCD had failed (by several orders of
magnitude) to give the correct result. The main uncertainties
involved in this estimate come from
the determination of the strength of the instanton effects at the
scale considered here, where, even though nonperturbative effects cannot
be neglected, one is already well within the perturbative region. Since
these effects can be brought under theoretical control, this provides a
strong motivation for  a systematic investigation of instanton-induced
helicity effects at medium-high energy of which Ref.~\anfor\
(for forward elastic scattering) and the present work are the first
steps.

\bigskip
\noindent{\bf Acknowledgements:} We thank L.~Magnea and C.~Rossetti
for discussions.

\vfill
\eject
\bigskip
\listrefs
\vfill
\eject
\centerline{\bf FIGURE CAPTION}
The diagram which leads to the decay  $\eta_c\to p\bar p$. The blob indicates
the effective instanton-induced interaction of Eq.~\effglu.
\bigskip
\bye